# Refuting Samuelson's Capitulation on the Reswitching of Techniques in the Cambridge Capital Controversy

*By* CARLO MILANA*

Birkbeck College, University of London

Correspondence: Carlo Milana, Department of Management, Birkbeck College, Malet Street, Bloomsbury, London WC1E 7HX. Email: c.milana@bbk.ac.uk

December 2019

**Abstract**: Paul A. Samuelson's (1966) capitulation during the so-called Cambridge controversy on the phenomenon of reswitching of techniques in capital theory had implications not only in pointing at a supposed internal contradiction of the marginal theory of production and distribution but also in preserving vested interests in the academic and political world to this day. Based on a new non-switching theorem, the present paper argues that Samuelson's capitulation was logically groundless from the point of view of the economic theory of production.

**Introduction**

In a recent article on the Cambridge (UK)-Cambridge (MA, USA) controversy in capital theory, the author of this paper conclusively proved a new non-switching theorem stating that *consistently with the marginalist theory of production and distribution and contrary to the conclusions of that controversy the so-called paradoxical "reswitching of techniques"* *appearing in the real wage-interest coordinate space can never occur in the coordinate space of true real factor-prices with any version of the Sraffian model,* In short, what you see is **not** what you get in the reswitching paradox. The proof of this theorem was substantiated with further evidence by revealing unexplored contradictory aspects of various numerical examples that were presented in the literature in support to the Sraffian criticism (specifically, Bruno *et al.*, 1966, Garegnani, 1966, 1970, 1976, Sato, 1966, Leibman & Nell, 1977, and Perz, 1980). These discussions referred to Sraffa's (1960: Chaps I-V, XII) intersectoral model with one-period circulating capital where non-linear effects of interest rate on prices arise from intersectoral interdependencies in refutation of the purported proof of a false non-switching theorem proposed by Levhary (1965), one of Paul Samuelson's research associates. Those prompted Sraffian counterexamples brought about Levhary & Samuelson's (1966) recognition of the error and Samuelson (1966) capitulation on the reswitching paradox envisioned by the Cambridge (UK) contenders. Because of this capitulation, the paradox of cost minimization leading to the reswitching of techniques over a monotonic succession of the interest rate levels has been widely accepted among "the facts of life" as an internal contradiction of the neoclassical theory of production.[1] This article aims at demonstrating that Samuelson's capitulation appears logically groundless at the light of a new non-switching theorem.

Since then, it has become customary to represent the "reswitching" phenomenon graphically by plotting, for each technique, the wage-interest relation (more precisely the trade-off relation between the real wage rate relative to the output price and the interest rate deriving from the respective accounting equation of the cost of production. Each of two wage-interest curves associated respectively with the alternative techniques $α$ and $β$ determines the real wage for a given interest rate. By increasing the interest rate progressively, the cost-minimizing system switches from *α* to *β*, at a particular level of the interest rate, and then it switches back to *α,* at higher levels of the interest rate.

---

[1] Samuelson (1966: 583) plainly declared: ''If all this causes headaches for those nostalgic for the old time parables of neoclassical writing, we must remind ourselves that scholars are not born to live an easy existence. We must respect and appraise, the facts of life''.

The debate on such a phenomenon has been recently revived, especially regarding its economic significance and its relevance for aggregation in current macroeconomic analyses. This result is still seen to have implications for the micro-foundations of macroeconomic models (Baqaee and Farhi, 2018), where the Sraffian criticism pointing to the noted inconsistency with the marginalist theory is still considered valid. As will be demonstrated below in this paper, the Sraffian criticism mistakes the interest rate for the capital input price. This misconception leads to a wrong conclusion pointing to an apparent internal inconsistency of the marginalist theory. It fails to note that, in such a paradox, a monotonic succession of the interest rate levels corresponds to a succession in opposite directions of the real capital-input price. The technical reswitching *α-β-α* over the range of interest rate would correspond to a single switch *α-β* over the range of the *relative rental prices* of capital goods. Each technique brings about a linear relation between these real prices and the relative wage rate. The central proposition of the present paper is, therefore, that, *consistently with the marginalist theory of production and distribution, the reswitching of techniques can never occur in the coordinate space of the true real factor prices with any configuration of the Sraffian model*.

Sraffa (1960: Chap VI) had proposed a further model where the capital goods consist of "maturing" labour inputs at different periods, where the non-linear intertemporal effects of wage rate on prices. In such a model, current costs of production are the sum of present values of dated labour inputs whereby capital inputs correspond to dated labour costs. Following Pasinetti's (1966) numerical example based on the intertemporal Sraffian specification, Samuelson (1966) used the simplest Austrian case of intertemporal production costs to account for the reswitching phenomenon. He claimed that his example 'tells more simply the full story of the twenty-fifth and eighth degree polynomials of the Sraffa-Pasinetti example of reswitching' (*Ibidem*: 569). As hinted by Sraffa himself (*Ibid*, Par. 45) and also noticed by Bruno *et al*. (1966: 528-529), the intersectoral and intertemporal Sraffian cases could be integrated into a general model. Our non-switching theorem could as well apply to this general model. However, in order to focus on Samuelson's (1966) conclusion, the following discussion will use his numerical example.

**Samuelson's model**

Samuelson's (1966: 571-73) second numerical example considers the production of one unit of champagne[2], described as follows:

> To help economic intuition, suppose champagne is the end product of both [techniques] II*a* and II*b*. In II*a*, 7 units of labour make 1 unit of brandy in one period. Then 1 brandy ferments by itself into 1 unit of champagne in one more period. In II*b*, 2 units of labour make 1 grapejuice in one period. In one further period 1 grapejuice ripens by itself into 1 wine. Then 6 units of labour shaking 1 unit of wine can in one more period produce 1 champagne. All champagne is interchangeable. (*Ibidem*: 571).

This example is illustrated in **Figure 1** below. It appears to be inspired by Wicksell's (1911: Vol. 1, 172-177) production case of "a certain kind of wine" where interest is seen as the marginal productivity of "waiting" (*ibidem*: 177). Samuelson aims at comparing the cost-effectiveness of two different methods of production over a wide range of the interest rate. The technique *b* dominates technique *a* with a lower cost when the interest rate level is fixed in the interval between 50 and 100 percent, whereas the technique *a* produces at a lower cost than technique *b* when interest is fixed below or above that interval of rates.

---

[2] The first numerical example proposed by Samuelson is not considered here as it does not lead to reswitching.

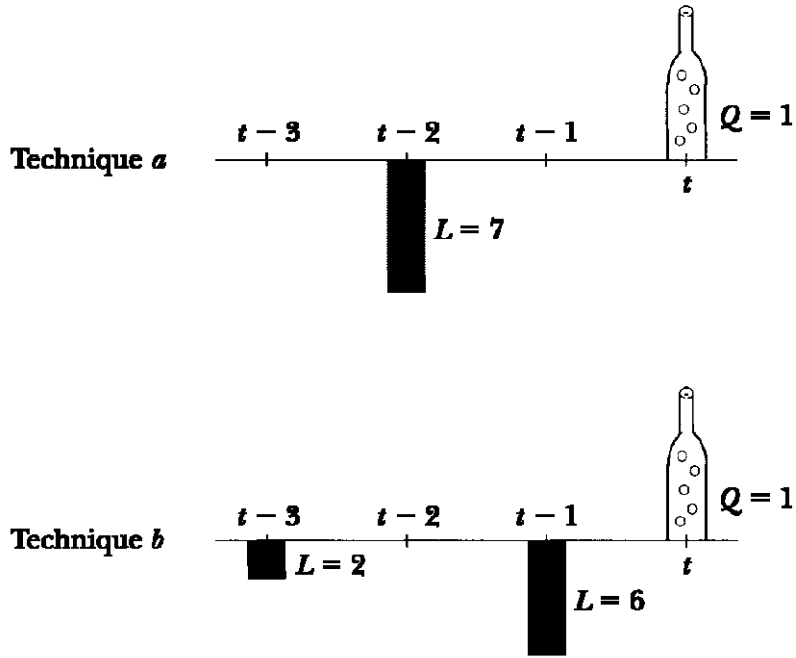

**FIGURE 1. SAMUELSON'S (1966: 571-573) SIMPLE AUSTRIAN MODEL** (in the graphical illustration drawn by Coen and Harcourt, 2003: 203). The two alternative techniques II*a* and II*b* have labour inputs {0, 7, 0} and {6, 0, 2} respectively. Technique *a* requires seven units of labour two periods before the bottle of champagne is a finished product and no labour in other periods, whereas technique *b* requires two units of labour three periods before and six units of labour one period before the champagne is made.

The original economic accounting of this model is expressed in formula (1) with the cost $c_h(w,i)$ of one bottle of champagne taking shape at the end of the third period with two alternative techniques *a* and *b* for given levels of a predetermined interest rate *i*:

(1) $$c_h(w,i) = w(1+i)L_{h1} + w(1+i)^2 L_{h2} + w(1+i)^3 L_{h3} \qquad \text{for } h = a,b$$

whereby *w* is the anticipated labour wage (paid at beginning of period), $L_{ht}$ is the labour service employed with technique $h = a,b$. The numerical values of parameters are *ante-factum wage* $w = \$1$

$$L_{a1} = 0; \quad L_{b1} = 6$$
$$L_{a2} = 7; \quad L_{b2} = 0$$
$$L_{a3} = 0; \quad L_{b3} = 2$$

The two alternative cost equations are therefore numerically defined:

$$c_a(w,i) = 7(1+i)^2$$
$$c_b(w,i) = 6(1+i) + 2(1+i)^3$$

**Samuelson's analysis**

**Table 1** reports the total costs for given levels of interest rate in the range of 150-0 per cent.

**TABLE 1.** COST OF ONE UNIT OF CHAMPAGNE

| Interest rate (percent) | Total cost w technique $a$ | Total cost wit technique $b$ (( |
|---|---|---|
| 150 | 43.75 | 46.25 |
| 125 | 35.44 | 36.28 |
| **100*** | **28.00*** | **28.00*** |
| 75 | 21.44 | 21.22 |
| **50*** | **15.75*** | **15.75*** |
| 25 | 10.94 | 11.41 |
| 0 | 7.00 | 8.00 |

*Switch point

The two techniques are 'tied' at the switching point $1+ i = 1 + 1.0$ corresponding to 100 percent interest per period with a total cost of $ 28 each. They are again 'tied' at $1 + i = 1 + 0.5$ corresponding to 50 percent interest per period with a total cost of $ 15.75 each. Technique II*a* dominates technique *b* in terms of cheaper cost at the interest rate higher than 100 percent and lower than 50 percent, whereas technique II*b* is cheaper between the two switch points at the intermediate levels of interest rate between 100 and 50 percent. This recurrence of a technique to dominance called 'reswitching' would appear as a paradox in contradiction with the technology convexity and the neoclassical monotonic function of input demand with respect to its own price. The paradox would then be in evident contrast with the marginal productivity theory according to which additional quantitative units of a factor of production are paid their monotonically decreasing marginal product.

Samuelson (1966: 572) provided **Figure 2** to summarize 'the effect of interest rate changes on the relative costs of producing champagne by the two method'**.**

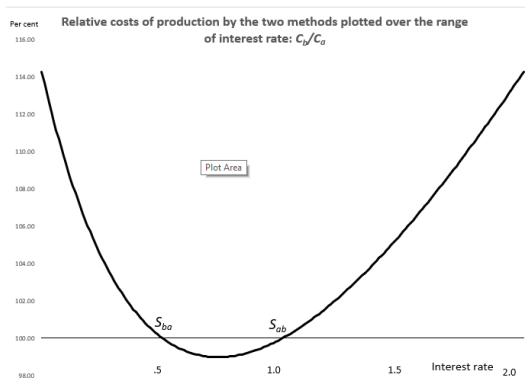

**FIGURE 2. RESWITCHING BACK TO II*a* FROM II*b*.**
It occurs because the plotted curve of relative costs is not one-directional over the range of interest rate. Between switch points, technique II*b* is used; elsewhere II*a* is used. Because of the non-linear effects of interest rate on relative prices, the cost ratio in the plotted

> curve occurs at two different interest rates, except in its minimum point between $S_{ba}$ and $S_{ab}$ corresponding to one single level of the interest rate. In general, there is no univocal relation between the relative costs and the interest rate.

The conclusion is that "[t]he fact of possible reswitching teaches us to suspect the simplest neoclassical parables" (*Ibidem*: 574.) Nine years later, Samuelson (1975) confirmed the point of view expressed in his 1966 summary:

> [H]ow much of my substantive argument evaporates, or is vitiated, or needs emendation and elucidation? None that I can see. No diagram needs redrawing. No substantive contention need be withdrawn or qualified (p. 44).

However, he immediately recognized that "it is only too easy to be blind to one's own shortcomings." While honoring his intellectual awareness, the following critical discussion is proposed.

**A critical discussion of Samuelson's interpretation**

In this numerical example, each level of the cost ratio occurs at two different levels of the interest rate, except in its minimum, as it is well known to financial engineers since a long time (Osborne and Davidson, 2016 provided a fresh view on reswitching arising from the multiple levels of interest rate in the choice of production techniques). By contrast, as it will be shown below in **Figure 3**, each level of the same cost ratio has a univocal relation with relative factor prices in full consistency with the neoclassical production theory. Samuelson (1966: 571, fn. 3) appealed to the authority of "John Hicks, *Capital and Growth* (Oxford: Clarendon Press, 1965), Chap. XIII, [where he] is in agreement with the upshot of the present symposium." However, in the previous Chap. XII of the same text, Hicks presented an intersectoral neoclassical model and expressed perplexity on the concept of interest rate as a factor price:

Samuelson, I fancy, would call it the 'factor-price equation' […] I do not, however, like the idea of regarding $i$, as well as [real wage] $w/p$, as a factor price. It is the [rental of the capital good] $r$ which is the factor price to my way of thinking. (Hicks, 1965: 140, fn. 1, Notation adjusted.)

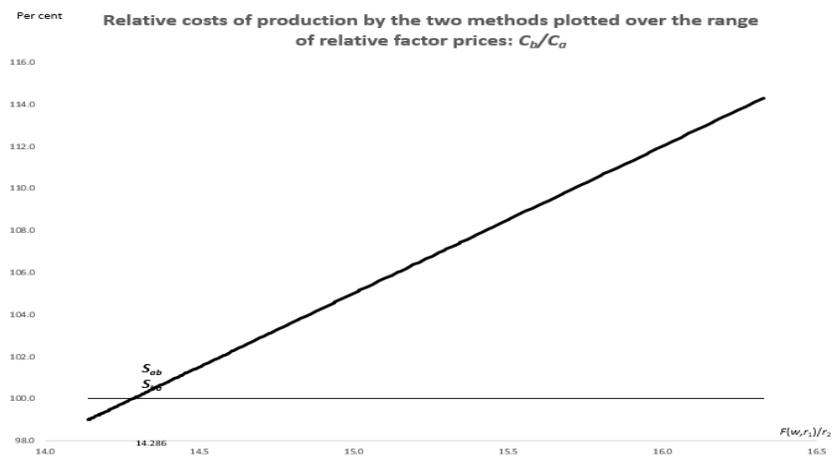

**FIGURE 3**. **Plotting the relative costs of production over the range of *relative factor prices*, only one single switch point is observed**, which is, however, consistent with *both* switch points $S_{ab}$ and $S_{ba}$ in **Figure 2**. The contrast between **Figures 2 and 3** is only superficial. Multiple interest rates solving project financing problems are well-known in financial engineering at least since Irving Fisher, recognized by Samuelson (1966: 581, fn. 2) himself as the 'multiplicity of Fisherian yield known ever since the 1930's.' Nevertheless, he failed to see that such multiplicity of yields is in no contradiction with the convex technology and marginal productivity theory as implied in this monotonic relation with relative factor prices.

Later, Hicks (1973: 39, fn. 2) criticized the interpretation of the wage-interest trade-off frontier again:

> It is still less appropriate, on the present approach, to describe it as a 'factor-price curve' or 'factor-price frontier'. In an 'Austrian' theory, *the rate of interest is not the price of a 'factor.* (Emphasis added.)

Model (1) can be easily reformulated in cost accounting in terms of *post-factum* wage $w_L$ and neoclassical rental prices $R_1$ and $R_2$ of capital goods as follows:

(2) $$p_h(w, r_1, r_2) = w_L a_{Lh} + R_1 a_{K_1 h} + R_2 a_{K_2 h} \quad \text{for } h = a, b$$

where

$$w_L = w(1+i)$$
$$R_1 = p_{K_1}(1+i) = (1+i)^2; \text{ where } p_{K_1} = (1+i)$$
$$R_2 = p_{K_2}(1+i) = (1+i)^3; \text{ where } p_{K_2} = (1+i)^2$$

with input-output coefficients defined as follows

$$a_{La} = L_{a1} = 0; \quad a_{Lb} = L_{b1} = 6$$
$$a_{K_1 a} = L_{a2} = 7; \quad a_{K_1 b} = L_{b2} = 0$$
$$a_{K_2 a} = L_{a3} = 0; \quad a_{K_2 b} = L_{b3} = 2$$

It can be noted, *in passim*, that the rental prices $R_1$ and $R_2$ are implied in Wicksell's (1911: 154) theory of interest paid by a portion of "marginal productivity of capital goods as saved up labour and land in the past. In his own words,

> The "*marginal* productivity of saved resources of labour and land is greater than that of the current resources. […] This marginal productivity, and the share in the product which it determines, provides in the first place, a recompense for the actual capital used in production, but it also provides something more. […] This surplus is what is called interest. We thus arrive at the following definition:—
> 
> *Capital is saved up labour and saved up land. Interest is the difference between marginal productivity of saved up labour and land and of current labour and land.* (Wicksell, 1911, Vol. 1: 153-154, Emphasis in original.)

In other words, interest is a component of the complex price called "rental price" or "user cost" of a saved resource, which, in turn, is compensated for the value of its marginal productivity. Therefore, the interest component of the rental price corresponds to the value of *net* marginal productivity, which is set aside to compensate lenders of capital for "waiting time." It should also be noted that the cost of production defined using the structural form (2) is identically equal to the cost defined using the reduced form (1), that is, in stationary equilibrium,

$$p_h(w_L, R_1, R_2) \equiv c_h(w, i)$$

*The model is characterized by three factors of production and two alternative techniques, featuring more inputs than techniques as a necessary condition for reswitching to appear in the wage–interest coordinate space.* The two techniques can now be compared in terms of cost ratio with reference to relative factor prices. **Table 2** reports the relative factor prices and the ratio of total costs with the two alternative techniques for given levels of the interest rate. Here the aggregate price $F(w_L, R_2)$ exists according to the Leontief (1947a, 1947b)-Sono (1945, 1961) condition on the existence of its dual quantity aggregate of two inputs $X(a_{Lb}, a_{K2b})$ since the price of another input does not affect their quantity ratio, that is $\dfrac{\partial}{\partial R_1} \dfrac{a_{Lb}}{a_{K2b}} = 0$.

**TABLE 2.** THE COST RATIO OVER THE RANGE OF RELATIVE FACTOR PRICES

| Relative factor price $F(w_L, R_2)/R_1$ | Rate of interest $i$ (percent) | $C_b/C_a$ (percent) |
|---|---|---|
| 6.9282* | 73.21* | 98.97* |
| **7.00**** | **50 and 100**** | **100.00**** |
| 7.17 | 33 and 125 | 102.38 |
| 7.30 | 25 and 140 | 104.29 |
| 7.40 | 20 and 150 | 105.71 |
| 8.00 | 0 and 200 | 114.29 |

\* Maximum cost advantage of technique *b*.
\*\*Unique switch point in the dimensional space of factor prices
Note: The price aggregate $F(w, R_2)$ is identically equal to $C_b$ in Samuelson's (1966) numerical example.

The focus of **Table 2** is that technique *b,* which is intensive in labour *and* capital good 2, is dominant when their aggregated factor price relative to the rental of capital good 1 is lower than that in the switch point. It is instead dominated by technique *a*, which is intensive in capital good 1, when the rental price of the capital good 1 is lower than that in the switch point. Such a result is entirely consistent with the marginalist production theory, disproving the erroneous Samuelson's (1966) interpretation overlooking the key correspondence between a unique vector of real factor prices and multiple levels of the interest rate.[3] By contrast, only one switch point occurs over the entire range of relative factor prices. Except for the point where the cost ratio reaches its minimum of 0.9897 at an interest rate of 73.21 percent, every other single level of relative factor price corresponds to two distinct levels of the interest rate.

This example demonstrates that, when both techniques are tied with the same relative factor prices, the cost ratio is equal to 1 and coincides with *all* the Sraffian switching points simultaneously if also the interest rate coincides. By converse, the unique solution in terms of relative prices generally corresponds to multiple values of interest rate and real wage as the two types of solutions are mutually consistent with multiple roots of the interest rate. This is the mathematical reason why multiple switch levels of *i* correspond to the same *unique* vector of relative factor prices to which all Sraffian switch points recur. From an economic point of view, the significance of such a result is that it would be in contrast with the purported *marginal productivity theory of interest* implied by the British Cantabrigeans with their criticism based on the "reswitching paradox." In plain words, the plural Sraffian switch points $S_{ab}$ and $S_{ba}$ occurring over the range of interest rate shown in **Table 1** and **Figure 2** correspond to a singular point in the coordinate space of relative factor prices shown in **Table 2** and **Figure 3**. The following theorem has been proved in the discussion above:

---

[3] The persistence of such erroneous interpretation of reswitching appearing over the range of interest rate is testified by recent studies such as Baqaee and Farhi (2018: 36-40).

**THEOREM. Impossibility of the Sraffian reswitching of techniques (Milana, 2019: 115)**. *The Sraffian reswitching of techniques of production that appears in the real wage-interest rate space never occurs in the corresponding space of real factor prices in full consistency with the marginalist theory of factor rewards.*

To be sure, influential authors soon expressed perplexities on the Sraffian approach. Following previous contributions, in a later article entitled 'Is Interest the Price of a Factor of Production?', Hicks (1979) came back to his previous question with an extended discussion on why financial capital, of which the rate of interest is the 'price', cannot be considered as a factor of production. He replied negatively to the question of whether an additional unit of *financial* capital directly yields an additional change in the volume of production just because it has to be invested in anticipated wages of labour or capital goods to affect, on the supply side, *their own marginal products*. One can also recall previous contributions dating back not only to Wicksell (1911) but also Metzler's *'The Rate of Interest and Marginal Product of Capital'* (1950, 1951) and its later discussants starting from Lerner (1953), who clarified the relations between the rate of interest and marginal private and social products of capital through non-financial capital.

Samuelson (1966: 580), on the other hand, had to appeal to elements outside the production technology to define the "net productivity of capital":

> [T]he present discounted value of all the *net* gains in future consumptions resulting from a switch from a process like *Ia* to a process like *Ib* (net in the sense that all sacrifices of consumption must be taken into account, with their proper discount, as subtractions) will balance out to zero, if the interest rate used in discounting is that of the switch point $S_{ab}$ between process I*a* and I*b*. In a bookkeeping sense, therefore, the Fisherian yield of product received in

> terms of product sacrificed is precisely measured by the market rate of interest.
> In this sense, the market rate measures the 'net productivity of capital'.

Whatever the semantics of the expression used, the term "marginal productivity of capital" in this context loses its direct connection with the technological convexity assumed in the neoclassical production function and its dual factor-price frontier. Instead, it is precisely against this very technological convexity that the purported internal inconsistency with the marginalist theory of the recurrence of techniques has been contrasted. The flawed interpretation of reswitching stems firmly from missing to observe the crucial point that *the multiple* switch points in the coordinate space of real wage-interest rate coincide with a unique switch point in the multidimensional coordinate space of real factor prices, *where reswitching can never appear*.

The effects of the interest rate on relative prices are well known to financial engineers at least since Irving Fisher (1907, pp. 352-53; 1930, p.279) and Hayek (1931, 1941). Fisher, in his classic works on interest, was also aware of the possibility of reversing capital *value* in relation to interest (Samuelson, 1966, p. 581, fn. 2, Velupillai, 1975, 1995). A series of critical contributions (Yeager, 1976, 1979; Garrison, 1979, 2006; Greenfield, 2003; Choi, 2017; Lewin & Cachanosky, 2019: 67-74) have already clarified that Samuelson's calculation is nothing but an investment evaluation, not a production decision implying no conflict between the multiple roots of the financial equations and the single roots of the factor demands derived from the well-behaved technology.

**Further considerations**

Samuelson (1966) failed to notice also another important phenomenon that was present in his numerical example: *complementarity*, which is a well-known concept opposed to substitutability established in the seminal paper by Hicks and Allen (1934), revisited by Samuelson (1974) himself as a tribute to those authors on the 40$^{th}$ anniversary of their celebrated

"revolution" in demand theory. Samuelson opened his important essay starting with "what every school boy knows":

> As Ludwig Wittgenstein would say, we "know" that coffee and tea are "substitutes" because we can drink one or the other; in the same way, we know that tea and lemon are "complements," because tea with lemon makes up our desired brew. (*Ibidem*: 1255.)

However, as he warned, "[t]he simplest things are often the most complicated to understand fully." Complementarity can be considered as a theoretical possibility with perverse behavior of demand in one particular subspace of factor inputs over some range of input prices while preserving consistency with the global convexity of technology in the coordinate space of all inputs. "Perversity" in demand behavior means that the demanded quantity increases (decreases) if the own price increases (decreases). Although "perversity" does not necessarily lead to reswitching, reswitching implies "perversity" of demand in some region of the relevant range of factor prices (Hatta, 1976). This fact was recalled, *in passim,* by Solow (1975: 52, fn. 3) in a later stage of the debate on reswitching in the following text:

> [S]uppose that from study of the two-good case I had concluded that all commodities were substitutes in the Slutzky sense; and then I learned that as soon as there are three or more goods complementarity is possible. I could now repeat the last sentence of the text.

which was the following:

> I would have to kiss a neat generalization good-by, and its immediate consequences too, but the theory of consumer demand would evidently not tumble on that account. (*Ibidem*: 52.)

It was soon recognized, for example, by Lachmann (1947), that also the theory of capital would not tumble in the case of a complex intersectoral economy with many complementary capital goods. The irony of this conclusion is that this type of reswitching arises from complementarity in the coordinate sub-spaces of individual *real factor prices,* whereas the first type of reswitching (the Sraffian type), due to the multiplicity of roots for interest, is instead confined in the coordinate *wage-interest* space.

Specifically, in Samuelson's (1966) second numerical example, the first and third factor of production (labour and capital goods no. 2, respectively) have been indeed found to be complementary by Hatta (1976: 130) in the following demonstration:

> Samuelson's perversely behaved production function consists of two techniques (*a*) and (*b*); (*a*)'s input vector ($x_1$, $x_2$, $x_3$) for producing a unit output is (0, 7, 0) and (*b*)'s (6, 0, 2). It can be shown that pair (1, 3) of this production function is complementary. Assume that under a certain structure of input prices, (*b*) costs less than (*a*) for producing the same amount of output. Then (*b*) will be employed. Suppose $p_1$ is increased keeping $p_2$ and $p_3$ constant. Sooner or later, technique (*b*) will become more expensive than (*a*), and (*a*) will be employed. Technique (*a*) uses less $x_3$ than (*b*) does in order to produce the same amount of output. This means that the rise in $p_1$, causes a reduction in $x_3$; namely, pair (1, 3) is complementary. Thus Samuelson's discrete model satisfies our necessary condition for perversity obtained for the neoclassical production function, namely the existence of a complementary pair of inputs.

The "perverse" behavior found by Hatta (1976) in Samuelson's example is defined concerning the above formula (2) featuring *rental prices* of capital goods rather than *the interest rate* as the price of (financial) capital in the reduced form (1). From empirical evidence, Morrison and Berndt (1981), for example, have obtained numerous instances of labour-equipment complementarity and "perverse" behavior of input demand in the US manufacturing industries (although these authors missed noting them explicitly).

**Final remarks**

Samuelson's (1966) capitulation in the Cambridge capital controversy seems to confirm the fallacy of "implicit theorizing" noted by Leontief (1937) in "the logical pattern used by [U.K.]

Cambridge economists." Three Symposiums hosted by economic journals (*Review of Economic Studies*, June 1962; *Quarterly Journal of Economics*, November 1966; and *Journal of Economic Perspectives*, Winter 2003) and a vast literature over more than sixty years (Birner, 2002; Coen and Harcourt, 2003; Mata, 2004; Backhouse, 2014) have failed to reach a consensus on why the reswitching of techniques would derive from a convex technology. This state of things poses questions on the way science progresses.

Kuhn's *The Structure of Scientific Revolutions* (1970) views science as heavily influenced by nonrational procedures with new theories more complex than those they usurp but not always any closer to the truth. The late Ludwig Wittgenstein, inspired by his dialogues with Sraffa (*e.g.* Sen, 2003 and Sinha, 2009), in his posthumously published *Philosophical Investigations* (1953), described how paradigms with different semantics cannot have points of contact. The Austrian philosopher questioned the meaning of the "direct inspection of paradigms" in the absence of a competent body of rules. In the context of the present paper, the lack of a "competent body of rules" was the lack of consensus on concepts such as factor prices and real factor price frontier faced by producers.

Samuelson's (1966) *Summing Up* misled the economic profession into misconceptions just by mistaking the interest rate for a real factor price. However, the fallacy of an impossibility theorem does not imply that a better one would also be false. The present paper and Milana's (2019) article have conclusively demonstrated that, due to a mismatch between changes in the interest rate and real rentals, the "reswitching of techniques" is arithmetically impossible in the Sraffian space of true real factor prices. As it turns out, the reswitching paradox is not at all paradoxical. Sraffa's results are invariably consistent with the marginalist theory of production and distribution.